\documentclass[12pt,a4paper]{article}

\usepackage{newtxtext,newtxmath}
\usepackage[utf8]{inputenc}
\usepackage{authblk}
\usepackage{setspace}
\usepackage[margin=1.25in]{geometry}
\usepackage{graphicx}
\graphicspath{ {./figures/} }
\usepackage{subcaption}
\usepackage{amsmath}
\usepackage{float}

\usepackage[numbers,sort&compress]{natbib}

\bibpunct{(}{)}{,}{n}{}{--}

\usepackage{hyperref}

\hypersetup{
	colorlinks=true,
	linkcolor=black,
	citecolor=black,
	urlcolor=blue
}

\let\cite\citep

\title{\textbf{Sub-50 Picosecond exceptionally Bright Perovskite Scintillation by Unlocking Giant Oscillator Strength}}

\author[1]{Chuanwei Dai\textsuperscript{\ensuremath{\dagger}}}
\author[1]{Yunbiao Zhao\textsuperscript{\ensuremath{\dagger}}}
\author[2*]{Xiao Ouyang}
\author[1]{Huaqing Huang}
\author[1]{Yulan Liang}
\author[1]{Jiaqi Bai}
\author[1]{Yingjie Song}
\author[1]{Jianhan Sun}
\author[1]{Yiqun Duan}
\author[1]{Wenjun Ma}
\author[1]{Senlin Huang}
\author[1]{Shufeng Wang}
\author[1,3*]{Jianming Xue}
\author[4*]{Xiaoping Ouyang}
\affil[1]{\raggedright \normalfont\small State Key Laboratory of Nuclear Physics and Technology, School of Physics, Peking University; Beijing 100871, China}
\affil[2]{\raggedright \normalfont\small School of Physics and Astronomy, Beijing Normal University, Beijing, China}
\affil[3]{\raggedright \normalfont\small CAPT, HEDPS and IFSA, College of Engineering, Peking University, Beijing 10087, China}
\affil[4]{\raggedright \normalfont\small Northwest Institute of Nuclear Technology, Xi'an, China}
\affil[ ]{\vspace{0em} \normalfont\small *Corresponding author. Email: jmxue@pku.edu.cn; oyxp2003@aliyun.com; oyx16@bnu.edu.cn}
\date{}

\begin{document}

\maketitle

%%%%%% Abstract %%%%%%
% 手动书写摘要标题
\begin{center}
    \textbf{Abstract}
\end{center}

\vspace{0.5em} % 调节标题与正文的间距

\noindent Ultrafast scintillators are indispensable for precise timing in high-energy physics and medical diagnostics. Fundamentally constrained by the trade-off between emission rate and light yield, conventional scintillators remain kinetically trapped in the sub-nanosecond regime, failing to break 50-picosecond limit. Here, we demonstrate a strategy to bypass this limitation by harnessing the coherent radiative acceleration in weakly confined CsPbCl\textsubscript{3} perovskite nanocrystals to generate an ultrafast photon burst. This effect originates from the giant oscillator strength, which we unlock by suppressing exciton-phonon scattering at mild cryogenic temperatures. Consequently, our scintillator achieves an unprecedented dominant lifetime of 13.11~ps alongside a high light yield of 21,851 photons/MeV. The resulting prompt photon emission rate more than 100 times higher than that of state-of-the-art ultrafast scintillators. We validate this breakthrough in realistic detection scenarios, achieving a coincidence time resolution of 30.8 ~ps and accurately resolving 13.5~ps electron bunches and 16.6~ps single-shot gamma-ray pulses. Our findings establish a robust coherent framework for next-generation ultrafast scintillators, pushing extreme radiation diagnostics into the picosecond frontier.

%%%%%% Main Text %%%%%%
%A few notes on the main text:  
%Three heading levels are permitted. Only the headings listed are permitted as Level 1 headings. Authors are encouraged to indicate the level of each heading using a unique format. For example, 
%\textbf{LEVEL 1 IN BOLD CAPS}
%\textbf{Level 2 in bold}
%\textit{Level 3 in italics}

\section*{\textbf{INTRODUCTION}}

\noindent \hspace{1em} Ultrafast scintillators capable of precise timing are indispensable for advancing frontiers in high-energy physics and medical diagnostics \cite{RN1, RN2}.  Specifically, breaking the 50-picosecond scintillation lifetime limit is critical to suppress signal pile-up in high-luminosity colliders and significantly boost the signal to noise ratio in medical imaging~\cite{RN3}. For decades, the pursuit of such capabilities has explored diverse systems, ranging from cross-luminescence crystals~\cite{RN4,RN5} and direct-bandgap semiconductors ~\cite{RN6} to thermally quenched inorganics~\cite{RN7} and organic plastic scintillators~\cite{RN8}. However, existing materials remain fundamentally limited by the trade-off between emission rate and light yield. They are kinetically trapped in the nanosecond regime, leaving the sub-50 ps largely inaccessible and failing to bridge the gap between present capabilities and next-generation requirements.

\noindent \hspace{1em} All-inorganic lead-halide perovskite nanocrystals~\cite{RN9,RN10,RN11}, particularly CsPbCl\textsubscript{3}, offer a compelling route to bypass this trade-off. By forming collective transition dipoles with giant oscillator strength  through quantum confinement ~\cite{RN12}, these materials theoretically permit radiative decay on the timescale of tens of picoseconds~\cite{RN12,RN13} Furthermore, they uniquely combine the high stopping power and radiation hardness of inorganic crystals with the rapid emission kinetics typical of organic plastics. Theoretical models dictate that in the weak quantum confinement regime---where the nanocrystal size exceeds the exciton Bohr radius---excitons can coherently delocalize over a macroscopic crystal volume. This phenomenon, governed by giant oscillator strength (GOS), facilitates the formation of collective transition dipoles, theoretically permitting radiative decay on the timescale of tens of picoseconds without sacrificing intrinsic luminosity.

\noindent \hspace{1em} Despite this theoretical promise, intense Fröhlich interactions with longitudinal optical (LO) phonons at ambient temperatures induce rapid quantum decoherence ~\cite{RN14,RN15}, forcing the exciton wavefunctions to collapse and severely restricting the effective transition probability. In this study, we demonstrate that deep freezing the system to 80 K effectively cuts off these thermal perturbations and dominant scattering modes. This thermal reconfiguration successfully transitions the material into a macroscopic coherent state, fundamentally unlocking the GOS-accelerated coherent emission pathway ~\cite{RN12,RN13}. Driven by this mechanism, the weakly confined CsPbCl\textsubscript{3} nanocrystals undergo extreme kinetic acceleration, yielding a dominant radiative lifetime of 13.11 ps (accounting for 83.36\% of the total emission) alongside an exceptionally high light yield of 21,851 photons MeV$^{-1}$. This achieves an unprecedented prompt photon brilliance ($B_p$) of 1.7$\times$10$^6$ photons MeV$^{-1}$ ns$^{-1}$, outperforming state-of-the-art ultrafast scintillators by over two orders of magnitude~\cite{RN16,RN17}. We validated this kinetic capability in situ, achieving a system-level coincidence time resolution (CTR) of 30.8~ps, and successfully resolving the real-time temporal profiles of \textasciitilde{}13.5~ps continuous electron bunches and 16.6~ps single-shot gamma-ray pulses. Collectively, these validations demonstrate that the coherent scintillation mechanism effectively shifts the fundamental temporal bottleneck in extreme radiation detection from the physical limits of luminescent materials to the bandwidth ceiling of downstream readout electronics ~\cite{RN18}.

\section*{\textbf{R}\textbf{ESULTS AND DISCUSSION}}
\label{sec:results_and_discussion}

\subsection*{\textbf{Giant oscillator strength-driven sub-50 }\textbf{ps}\textbf{ coherent scintillation}}

\noindent \hspace{1em} To circumvent the kinetic bottleneck of incoherent spontaneous emission, we unlocked a macroscopic coherent radiation pathway by harnessing the giant oscillator strength (GOS) of weakly confined all-inorganic CsPbCl\textsubscript{3} perovskite nanocrystals (NCs). Transmission electron microscopy (TEM) analysis confirms that the synthesized NCs exhibit a well-defined cubic perovskite lattice and are highly monodisperse with an average diameter of 21.87 $\pm$ 2.39 nm (Fig. 1A). Crucially, this size is significantly larger than the exciton Bohr radius of CsPbCl\textsubscript{3} (\textasciitilde{}2.5 nm)~\cite{RN19}, placing the NCs within the weak quantum confinement regime. Unlike strong confinement, this regime facilitates the coherent delocalization of excitons over a larger crystal volume, thereby enhancing the oscillator strength---a key prerequisite for accelerating radiative decay~\cite{RN12,RN20,RN21}. Driven by the GOS effect, at a cryogenic temperature of 80 K, our CsPbCl\textsubscript{3} nanocrystal scintillator undergoes an extreme kinetic acceleration. Time-correlated single-photon counting (TCSPC) reveals an unprecedented dominant radiative lifetime ($\tau$\textsubscript{1}) of 13.11 ps (accounting for 83.36\% of the total emission) (Fig. 1B). 

\noindent \hspace{1em} To quantify how the extreme photon emission rates arising from such coherent emission translate into a practical detection advantage,we establish a new figure of merit, termed prompt photon brilliance ($B_p$), which is formally defined as the absolute light yield divided by the scintillation lifetime ($B_p$ =LY/$\tau$). As illustrated in the reconstructed performance benchmark map (Fig. 1C)---plotted against diagonal $B_p$ contour lines---conventional materials face a rigid physical dichotomy. Traditional inorganic crystals (e.g., LYSO:Ce~\cite{RN22}, BGO~\cite{RN23}) deliver high light yields but are kinetically slow (falling into the 10\textsuperscript{2}$-$10\textsuperscript{3} $B_p$ regime), whereas ultrafast organic plastics (e.g., EJ228)~\cite{RN24} achieve rapid decay at the severe expense of photon statistics. In comparison, our GOS-driven CsPbCl\textsubscript{3} system, simultaneously achieving a sub-50 ps extreme lifetime and an exceptionally high light yield of 21,851 photons MeV$^-1$, achieves a $B_p$ of 1.7$\times$10\textsuperscript{6} photons MeV$^{-1}$ ns$^{-1}$. This $B_p$ value positions the material in a distinct performance regime, exceeding current state-of-the-art ultrafast scintillators by over two orders of magnitude. Such exceptionally high $B_p$ values fundamentally suppress the timing jitter originating from photon-statistical fluctuations~\cite{RN25,RN26}. In high-energy physics (HEP) and burn diagnostics for inertial confinement fusion (ICF), this property acts as an intrinsic physical filter: by generating ultra-steep, Dirac-like pulses with extreme instantaneous signal to noise ratios, it provides a definitive solution to the event pile-up problem encountered in high-luminosity environments ~\cite{RN27}.

\begin{figure}[H]
  \includegraphics[width=1\textwidth]{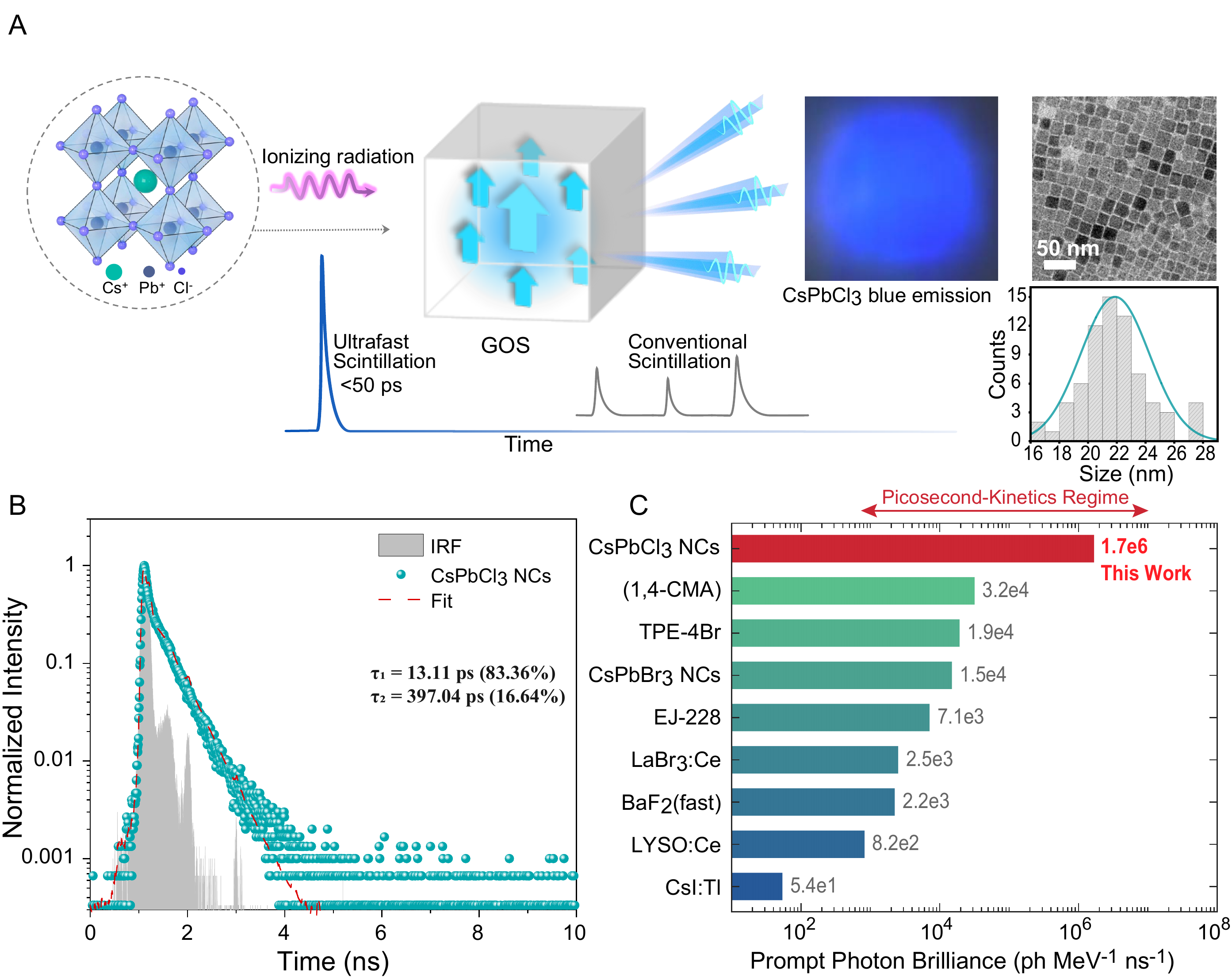}
\end{figure}

\begin{spacing}{1.0}
	\small
	\noindent\textbf{Fig. 1. Ultrafast scintillation from CsPbCl\textsubscript{3} NC scintillator.} (A) Schematic illustration of the GOS-enhanced coherent scintillation mechanism. Upon ionizing radiation, the coherent coupling of transition dipoles triggers superradiation (an ultrafast photon burst, $< 50~\mathrm{ps}$). The insets show the cubic perovskite crystal and a photograph of the scintillator exhibiting bright blue radioluminescence, and a representative transmission electron microscopy (TEM) image of the CsPbCl\textsubscript{3} nanocrystals. The inset histogram reveals an average diameter of 21.87 $\pm$ 2.39~nm, securely placing the system within a weak quantum confinement regime to provide the necessary physical volume for macroscopic exciton delocalization. Scale bar, 50~nm. (B) Extreme time-resolved photoluminescence (TRPL) decay profile at 80~K. The decay data is dominated by an unprecedented ultrafast lifetime component of $\tau_1 = 13.11~\mathrm{ps}$ (accounting for 83.36\% of the emission), fundamentally surpassing the 50~ps timing requirement. (C) Scintillator performance benchmark map charting light yield against scintillation lifetime, plotted with diagonal prompt photon brilliance ($B_p$) contour lines. Driven by GOS, CsPbCl\textsubscript{3} nanocrystal scintillator simultaneously achieves an unprecedented $B_p$ of $1.7\times10^{6}~\mathrm{photons\,MeV^{-1}\,ns^{-1}}$. This metric positions our material in a distinct performance regime (red bar), outperforming commercial inorganic crystals (such as LYSO:Ce , BaF\textsubscript{2}) and ultrafast organic plastics (such as EJ228) by over two orders of magnitude.
\end{spacing}

\subsection*{\textbf{Comprehensive scintillation performance and extreme radiation hardness}\textbf{ of CsPbCl}\textsubscript{3}\textbf{ NC scintillator}}

\noindent \hspace{1em} To comprehensively evaluate the scintillation performance of our CsPbCl\textsubscript{3} NC scintillator, we investigated the emission characteristics across different radiation environments. As shown in Fig. 2A, the CsPbCl\textsubscript{3} NCs exhibit highly consistent near-band-edge emission whether ionized by X-rays (radioluminescence, RL) excitation, high-flux electron beams (cathodoluminescence, CL), or optically pumped by a 375 nm laser (photoluminescence, PL). Compared to UV excitation, a distinct spectroscopic redshift (from 2.99 eV to 2.92 eV) emerges under ionizing radiation. This shift is a distinct signature of bandgap renormalization (BGR), where strong Coulombic interactions within the dense carrier within individual nanocrystals, lead to a downward shift of the band edge energy ($(\Delta E_{\text{BGR}} \propto n^{\alpha})$). Consequently, the emitted photon energy is reduced (E = E\textsubscript{g} -E\textsubscript{b}\textsubscript{ }- $\Delta$E\textsubscript{BGR}), providing direct spectral evidence that the ultrafast scintillation arises from a highly dense, interacting exciton formed within the ionization track~\cite{RN28,RN29}.

\noindent \hspace{1em} The macroscopic manifestation of the GOS effect is deeply rooted in the cryogenic suppression of decoherence pathway~\cite{RN30}. Temperature-dependent spectroscopic analysis reveals a dramatic convergence in the emission full-width at half-maximum (FWHM) as the system cools from 300 K to 80 K (Fig. 2D). This distinct spectral narrowing provides direct evidence for the suppression of thermal fluctuations and the effective cutoff of Fröhlich interactions (longitudinal optical (LO) phonon scattering). Consequently, as the exciton--phonon dephasing channel is progressively frozen out at cryogenic temperatures, long-range spatial coherence is established among the excitonic transition dipoles, enabling their constructive superposition into a macroscopic coherent dipole moment and thereby unlocking the full oscillator strength embodied in the GOS ~\cite{RN30,RN31}. This is macroscopically validated by the exponential increase in the absolute light yield upon cooling to 80 K (Fig. 2B). The simultaneous realization of high light yield and ultrafast response enables our material to redefine the limits of scintillator performance (Fig. 2C). This breakthrough completely breaks the long-standing trade-off between high-luminosity inorganic crystals (e.g., LYSO:Ce) with slow decay kinetics, and ultrafast scintillators (e.g., BaF2) constrained by low photon output.

\noindent \hspace{1em} Practical dosimetry in high-luminosity colliders and inertial confinement fusion strictly requires extreme radiation hardness. Under continuous high-flux electron beam bombardment, the GOS-enhanced perovskite scintillator exhibits a strictly linear dynamic response up to an extreme dose rate of $\sim 57\ \mathrm{kGy\ s^{-1}}$ (Fig.~2E), a broad operational window that fully satisfies the fluence requirements of advanced diagnostic scenarios. However, susceptibility to radiation-induced structural damage remains a primary limitation of ultrafast organic plastics. Owing to the intrinsic defect-tolerance~\cite{RN19} and robust ionic framework of the all-inorganic perovskite lattice, our material demonstrates extraordinary radiation hardness. Even after accumulating an absorbed dose of 40~MGy, the scintillator retains over 95\% of its pristine luminosity (Fig.~2F). This unparalleled radiation hardness far exceeds that of conventional ultrafast plastics (e.g., EJ-200)~\cite{RN32}, ensuring long-term operational reliability in extreme radiation fields. 

\begin{figure}[H]
  \centering
  \includegraphics[width=1\textwidth]{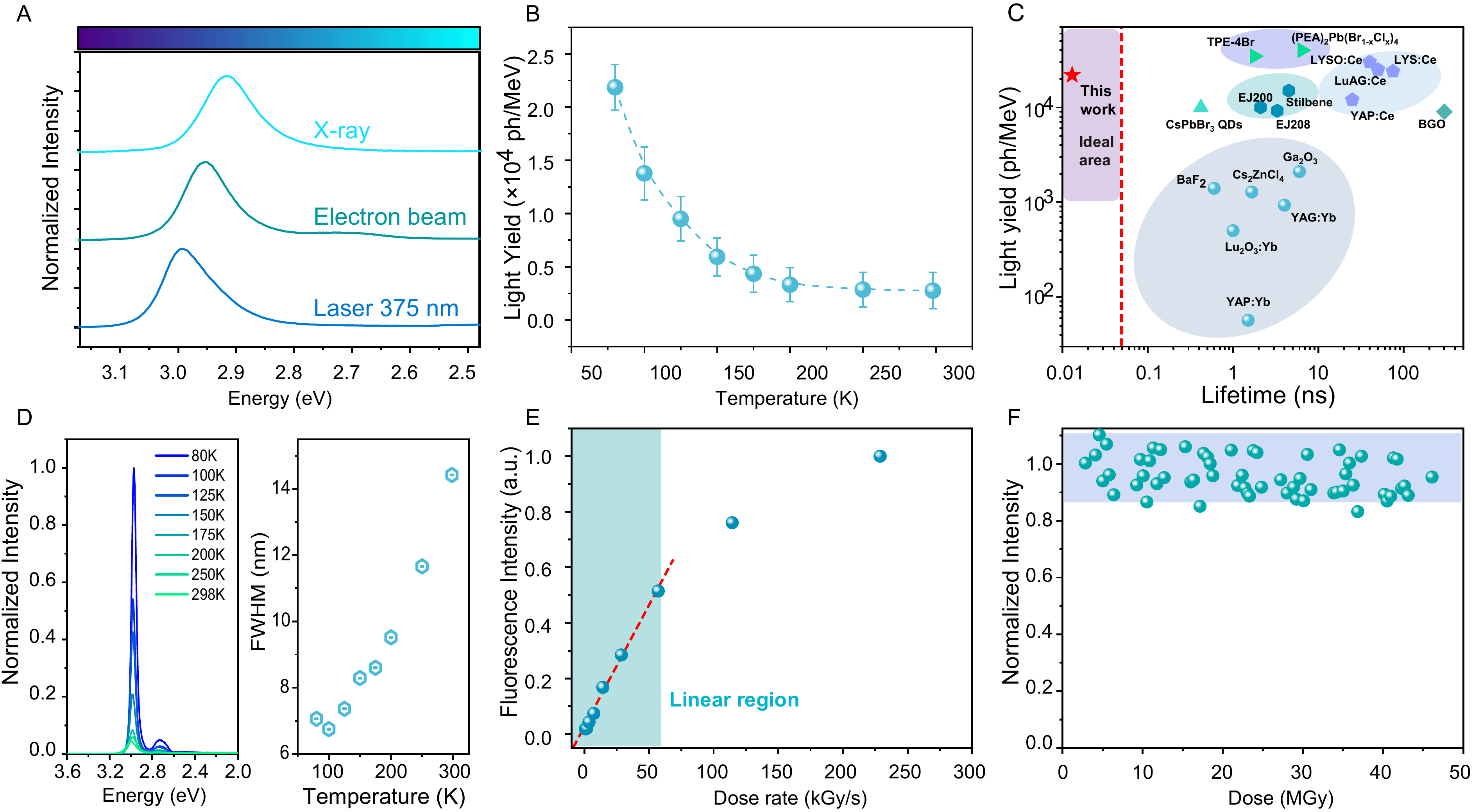}
\end{figure}  
  
\begin{spacing}{1.0}
\small
\noindent\textbf{Fig. 2. Comprehensive scintillation performance and extreme radiation hardness of the GOS-driven perovskite scintillator. }(A) Normalized emission spectra under continuous X-ray (radioluminescence, RL), high-energy electron beam (cathodoluminescence, CL), and 375-nm pulsed laser (photoluminescence, PL) excitation. The highly consistent near-band-edge emission demonstrates lossless thermalization of hot carriers. The distinct redshift under ionizing radiations (CL and RL) serves as a hallmark of bandgap renormalization (BGR) induced by an exceptionally dense exciton population within the ionization track. (B) Temperature-dependent absolute light yield from 300 K to 80 K. The exponential leap in photon yield at cryogenic temperatures macroscopically validates the full unlocking of the GOS effect. (C) Scintillation performance benchmark charting light yield against decay time. Driven by GOS, our CsPbCl\textsubscript{3} nanocrystal scintillator (red star) successfully occupies the ideal region (sub-50 ps and ultrahigh yield), fundamentally breaking the rigid speed-yield trade-off that constrains conventional inorganic and organic scintillators. (D) Evolution of the CL emission spectra (left) and the corresponding peak position and full-width at half-maximum (FWHM) (right) as a function of temperature. The sharp convergence of the FWHM provides direct spectroscopic evidence for the deep freezing of thermal fluctuations and the effective suppression of longitudinal optical (LO) phonon scattering. (E) Integrated scintillation intensity as a function of continuous high-flux electron beam dose rate. The broad and strictly linear response (shaded blue region) confirms its remarkable resistance to high-flux quenching, satisfying the demands of advanced dosimetry. (F) Extreme radiation hardness evaluation. The scintillator retains >95\% of its pristine radioluminescence intensity after enduring a cumulative absorbed dose of >40 MGy, demonstrating extraordinary intrinsic defect-tolerance superior to conventional ultrafast plastics.
\end{spacing}

\subsection*{\textbf{Temperature-driven exciton coherence and macroscopic GOS physical model}}

\noindent \hspace{1em} To elucidate the physical origin of the unprecedented coexistence of sub-50~ps extreme decay and ultrahigh light yield, we investigated the temperature-driven evolution of the exciton coherent state. In the CsPbCl$_3$ lattice at room temperature, intense Fröhlich interactions with longitudinal optical (LO) phonons induce rapid quantum decoherence~\cite{RN33}. This strong exciton-phonon scattering forces the exciton wavefunctions to collapse and localize within a restricted spatial volume, severely limiting the transition probability. However, as the system is cooled to 80~K, the thermal energy drops significantly below the LO phonon energy, effectively freezing the dominant scattering modes. Liberated from these thermal perturbations, the exciton coherence length $L_{\phi}$ extends to encompass the entire nanocrystal, enabling the transition dipoles of all unit cells within the coherent volume to superpose constructively~\cite{RN34}. This collective, in-phase coupling of $N_{\text{coh}} \propto (L/a_B)^3$ unit-cell dipoles constitutes the GOS effect, amplifying the single-exciton oscillator strength by a factor of $\approx (8/\pi)(L/a_{\text{B}})^3 \approx 1300$ relative to the strong-confinement limit (Fig. 3A)~\cite{RN35}.

\noindent \hspace{1em} Temperature-dependent time-resolved photoluminescence (TRPL) spectroscopy provides crucial dynamics for this mesoscopic quantum process. In conventional semiconductors, cooling typically prolongs the emission lifetime due to the thermodynamic freeze-out of non-radiative recombination channels~\cite{RN36,RN37,RN38,RN39}. In contrast, the CsPbCl\textsubscript{3} nanocrystals exhibit a dramatic acceleration of radiative recombination upon cooling (Fig.~3B): the dominant radiative lifetime decreases to 13.1~ps at 80~K, corresponding to a radiative rate $k_{\text{rad}} \approx 76~\text{ns}^{-1}$. Quantitative analysis reveals that the radiative rate increases by more than one order of magnitude---from $\sim 5~\text{ns}^{-1}$ at 200~K to $\sim 76~\text{ns}^{-1}$ at 80~K---following a strongly nonlinear temperature dependence (Fig.~3C). This behavior is quantitatively captured by our unified theoretical model. The close agreement between the theoretical curve and the experimentally extracted radiative rates confirms that the observed ultrafast emission originates from phonon-mediated modulation of the coherent oscillator strength, rather than from conventional nonradiative channel dynamics.

\noindent \hspace{1em} A size-dependent coherence scaling law unambiguously delineates the physical boundaries governing this extreme radiative performance (Fig.~3D). In the weak-confinement regime ($L \gg a_{\rm B}$), the exciton oscillator strength scales as $f \propto L^3$ due to the GOS effect---the coherent superposition of unit-cell dipoles across the nanocrystal's volume. However, two competing mechanisms ultimately arrest this cubic growth at larger sizes. First, electromagnetic retardation introduces destructive interference among dipoles separated by distances comparable to the emission wavelength: the retardation form factor $\mathcal{F}_{\rm ret} \approx \mathrm{sinc}^2 \left( \frac{\pi n_{\rm NC} L}{\lambda_{\rm exc}} \right)$ begins to suppress $f$ when $L \gtrsim \lambda_{\rm exc} / (\pi n_{\rm NC})$~\cite{RN40}. Second, spatial decoherence truncates the effective coherent volume to $L_\phi^3$ when $L$ exceeds the temperature-dependent coherence length $L_\phi$~\cite{RN30}, yielding a saturation factor $\mathcal{F}_{\rm coh} = 1 / [1 + (L / L_\phi)^3]$. The interplay of these three factors---GOS amplification, retardation suppression, and decoherence truncation---produces a well-defined optimal size $L_{\rm opt}$ at which the oscillator strength is maximized. At 80~K, where $L_\phi \approx 18$~nm, our model predicts $L_{\rm opt} \approx 20$--$30$~nm, in excellent agreement with the experimentally measured lifetimes across the available size range (Fig.~3D). This quantitative concordance validates the coherence scaling framework and reveals that the 20~nm nanocrystals used in this work operate near the theoretical optimum for radiative enhancement. Operating near this theoretical optimum ($L \approx 20$~nm), our nanocrystals simultaneously unlock the 13.11~ps extreme lifetime and record-high $B_p$. These findings establish a definitive size-coherence scaling framework for designing next-generation ps-level scintillators.

\begin{figure}[H]
  \includegraphics[width=1\textwidth]{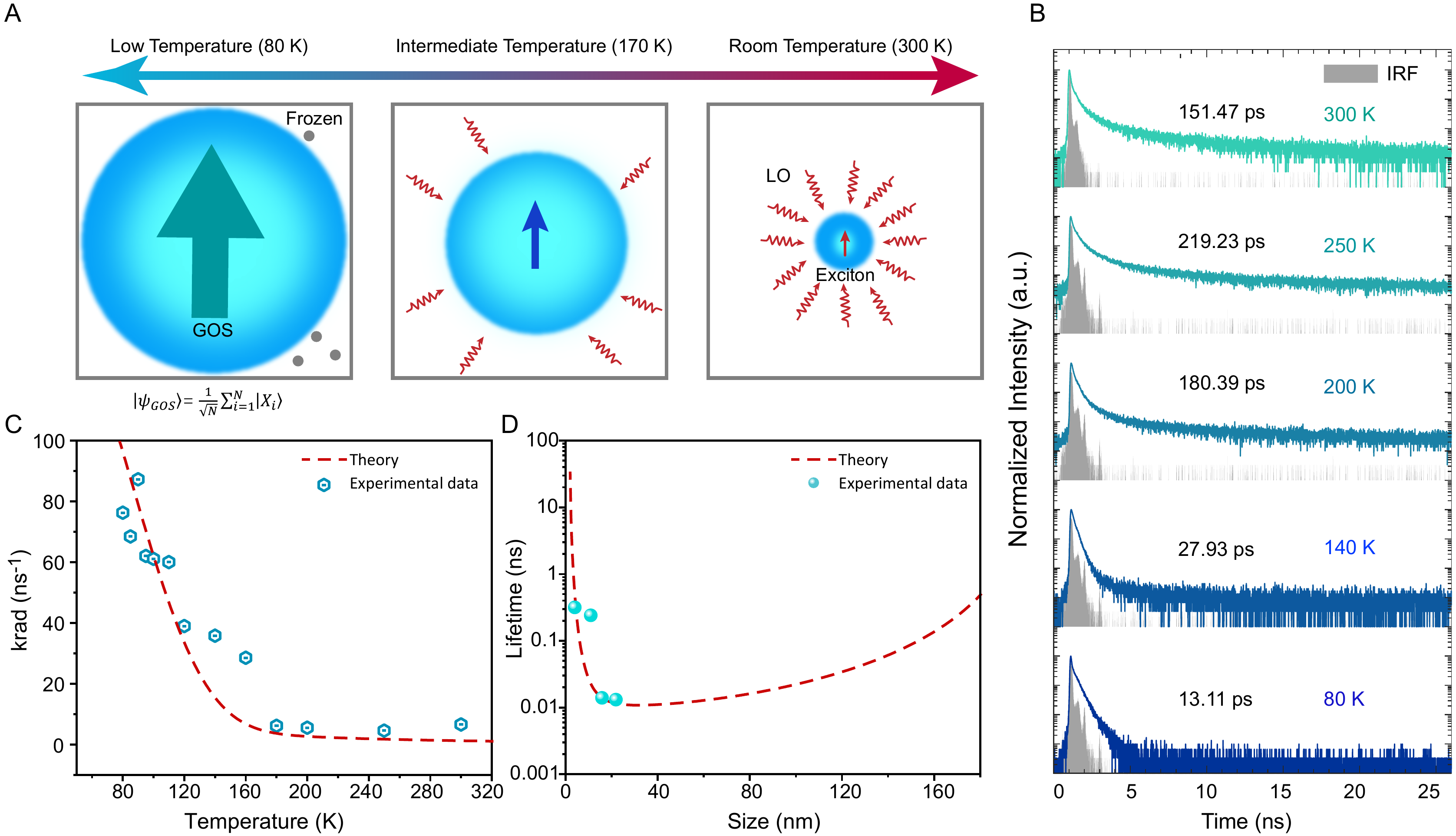}
\end{figure}

\begin{spacing}{1.0}
\small
\noindent\textbf{Fig.~3. Temperature-driven exciton coherence and the physical origin of GOS-accelerated scintillation.} (A)~Schematic illustration of the temperature-driven GOS coherent radiation model. At room temperature (300~K), intense Fröhlich interactions (longitudinal optical phonon scattering) severely restrict the exciton coherence length. Upon deep freezing at 80~K, thermal perturbations are effectively cut off, driving a non-linear expansion of the exciton coherent volume to encompass the entire nanocrystal volume. This thermal reconfiguration transitions the system into a macroscopic coherent state, formally unlocking the GOS-accelerated ultrafast emission. (B)~Temperature-dependent time-resolved photoluminescence (TRPL) spectra. In stark contrast to the conventional thermodynamic freeze-out behavior (lifetime prolongation) typical of standard semiconductors, the dominant radiative lifetime exhibits pronounced kinetic acceleration upon cooling, ultimately collapsing to the extreme 13.11 ps regime at 80~K. This serves as the direct dynamical evidence for the macroscopic GOS effect. (C)~Temperature dependence of the radiative rate $k_{\mathrm{rad}}$ for L = 20~nm CsPbCl\textsubscript{3} nanocrystals deposited on a quartz substrate. Open circles are experimental values extracted from TRPL decay analysis. The dashed red curve is the prediction of the unified theoretical model. Upon cooling from 300 to 80~K, k\textsubscript{rad} increases by more than one order of magnitude---reflecting the recovery of mesoscopic exciton coherence as phonon scattering channels are progressively frozen out. (D)~Size-dependent total lifetime of CsPbCl\textsubscript{3} nanocrystals at 80 K (log scale). Cyan circles denote experimental data; the dashed red curve is the theoretical prediction from the unified model. In the weak-confinement regime ($L \gg a_{\mathrm{B}}$), the lifetime initially decreases as $L^{-3}$ owing to the GOS effect, reaches a minimum of ${\sim}10$~ps near$L_{\mathrm{opt}} \approx 20\text{--}30$~nm, and then increases at larger sizes as electromagnetic retardation ($\mathcal{F}_{\mathrm{ret}}$) and spatial decoherence truncation ($\mathcal{F}_{\mathrm{coh}}$ progressively suppress the coherent oscillator strength. The non-monotonic behavior reveals an optimal nanocrystal size window that maximizes the radiative rate, within which the 20 nm crystals used in this study reside.
\end{spacing}

\subsection*{\textbf{Ultrafast extreme radiation diagnostics and temporal waveform resolution}}

\noindent \hspace{1em} To translate the GOS-driven coherent dynamics into practical detection breakthroughs, we benchmarked the scintillator for extreme high-precision radiation timing and real-time waveform characterization. In time-of-flight positron emission tomography (TOF-PET), minimizing timing jitter is critical. Using a standard \textsuperscript{22}Na positron source and a dual silicon photomultiplier (SiPM) coincidence setup (Fig. 4A), the CsPbCl\textsubscript{3} nanocrystal scintillator achieved a system-level coincidence time resolution (CTR) of 30.8 ps (Fig.~4B, C). Driven by the prompt photon emission and rapid rising edge, this precision approaches the intrinsic transit-time spread limits of current photoelectric readouts, establishing a viable pathway toward sub-millimeter spatial resolution in clinical PET diagnostics.

\noindent \hspace{1em} Beyond discrete event timing, we utilized the material for real-time continuous waveform characterization at the Peking University Superconducting Radio-Frequency (SRF) accelerator facility (Fig. 4D). Standard characterization of high-repetition-rate, ultra-short electron bunches relies on fast current transformers (FCTs). However, electronic bandwidth limitations in FCTs artificially broaden the \textasciitilde 10 ps intrinsic bunches into nanosecond-scale artifacts (measured FWHM $\approx$ 4.10 ns, Fig. 4E, F), obscuring the true temporal profile. By integrating the nanocrystal scintillator screen directly into the SRF beamline, we acquired halo-free, high-fidelity spatial profiles that precisely replicate the beam's intrinsic spatial distribution (Fig. 4G, H). Given its picosecond-scale response, we deconvolved the scintillator signal to extract the true temporal profile of the electron bunch. The reconstructed waveform exhibits an FWHM of 13.5 ps (Fig. 4I), in striking agreement with theoretical beam dynamics. By outperforming established commercial scintillators - including BaF\textsubscript{2} (0.64 ns) (S18), LYSO (28.34~ns) (S19), and YAP:Ce (36.61 ns) (S20), establishing our CsPbCl\textsubscript{3} nanocrystal scintillator as the benchmark for extreme fast-timing diagnostics.

\noindent \hspace{1em} Finally, to validate temporal resolution under extreme conditions, we measured single-shot, high-energy neutral photons---specifically, gamma-ray pulses generated via inverse Compton scattering (ICS) (Fig. 4J). Unlike charged particles, characterizing ultra-short neutral photon pulses is inherently challenging due to the inadequate response times of conventional high-density scintillators. The GOS-driven detector successfully resolved the ultrafast gamma pulses, yielding a deconvolved intrinsic FWHM of 14.2 $\pm$ 1.4 ps with a near-limit rise time of \textasciitilde{}1.0 ps (Fig. 4K). Collectively, these in situ measurements demonstrate that the coherent scintillation mechanism effectively shifts the fundamental bottleneck in extreme radiation detection from the physical limits of the luminescent material to the bandwidth ceiling of downstream readout electronics.

\begin{figure}[H]
\includegraphics[width=1\textwidth]{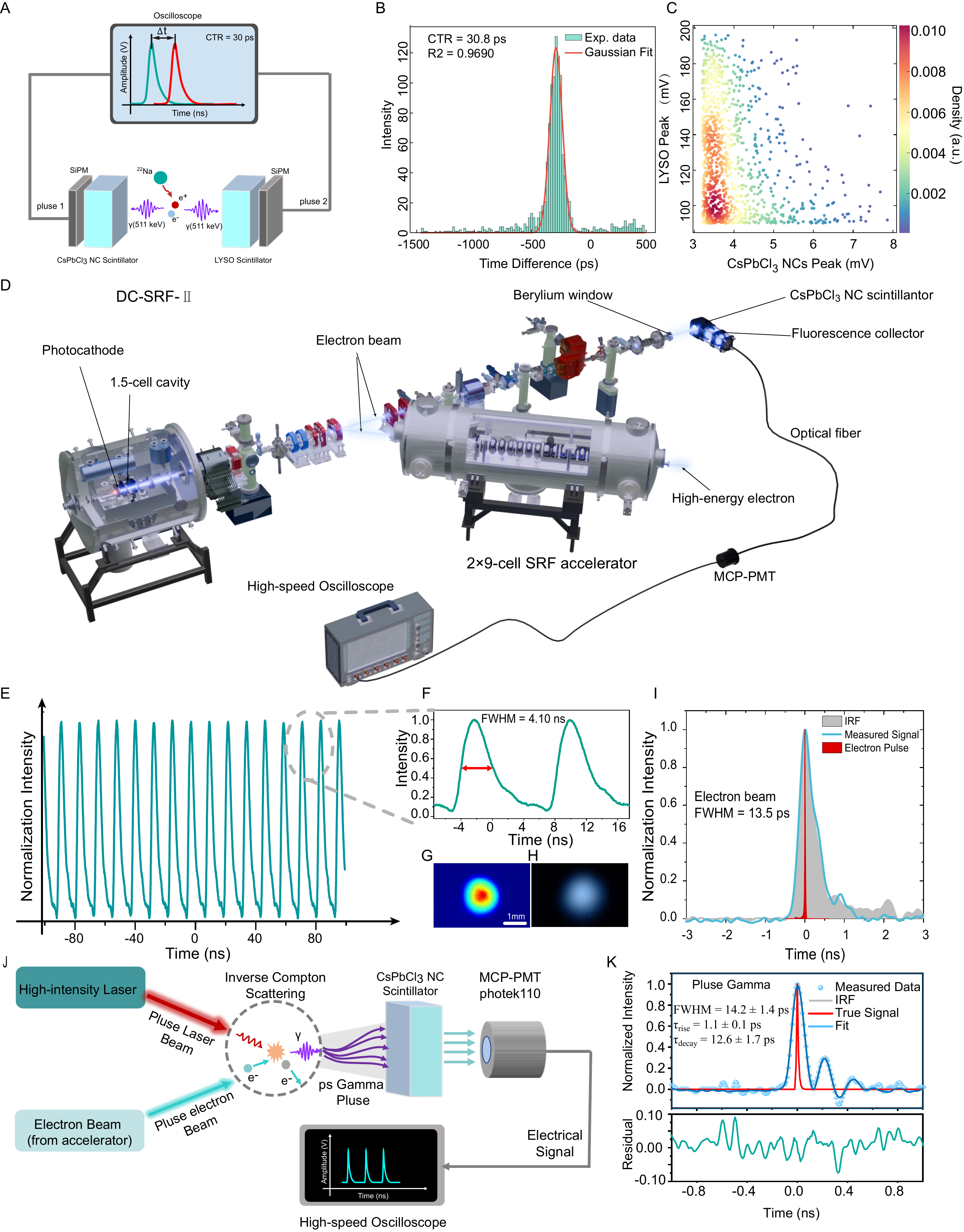}
\end{figure}

\begin{spacing}{1.0}
	\small
	\noindent\textbf{Fig.~4. Breaking the limit: Direct diagnostics of 10 ps ionizing radiation pulses.}(A) Experimental setup for high-precision temporal diagnostics. A Schematic of the coincidence time resolution (CTR) measurement setup employing a \textsuperscript{22}Na positron source and dual silicon photomultiplier(SiPM) arrays. (B) Coincidence spectrum for CsPbCl\textsubscript{3} nanocrystal scintillators with a Gaussian fit, demonstrating a system-level coincidence time resolution (CTR) of 30.8~ps. (C) Two-dimensional scatter plot of coincidence time versus signal amplitude (energy) for an asymmetric detection architecture, with a standard LYSO:Ce crystal on one side and a CsPbCl\textsubscript{3} nanocrystal scintillator on the other. (D) Application in superconducting accelerator beam diagnostics. Three-dimensional rendering of the superconducting radio-frequency (SRF) accelerator test facility at Peking University. While generating ultra-short high-energy electron bunches (theoretical pulse width \textasciitilde{}10 ps), standard fast current transformers (FCTs) are limited by nanosecond-scale response times, preventing accurate resolution of the true waveform. (E) Continuous macro-pulse waveform measured by standard FCT (conventional diagnostic method). (F) Magnified view of a typical FCT-measured electron bunch pulse, showing severe signal broadening (FWHM $\approx$ 4.10 ns). (G) Intrinsic spatial distribution of the electron beam spot. (H) High-fidelity luminescent spot from the CsPbCl\textsubscript{3} nanocrystal scintillator screen under pulsed electron beam excitation, precisely replicating the beam's spatial profile. (I) Temporally resolved profile from the CsPbCl\textsubscript{3} scintillator. Leveraging the material's inherent picosecond-scale ultra-fast response and deconvolution, the true electron pulse waveform (red filled curve) is successfully extracted, yielding an FWHM of 13.5 ps, closely matching theoretical predictions and bridging the gap between theoretical beam parameters and actual diagnostic capabilities. (J)Schematic of the experimental setup for generating and directly diagnosing ultrafast gamma-ray pulses via inverse Compton scattering (ICS). (K) Single-shot temporally resolved profile of the ICS gamma-ray pulse captured by the CsPbCl\textsubscript{3} nanocrystal scintillator detector. Deconvolved intrinsic gamma-ray pulse (green dashed line) exhibits an FWHM of 14.2 $\pm$ 1.4 ps and a near-limit rise time of \textasciitilde{}1.0 ps, successfully enabling precise dynamic tracking of extremely short high-energy photon bunches.
\end{spacing}

\section*{\textbf{CONCLUSION }}

\noindent \hspace{1em} In summary, we have demonstrated that unlocking the macroscopic giant oscillator strength (GOS) in weakly confined CsPbCl\textsubscript{3} nanocrystals provides a fundamental solution to the historical speed-yield trade-off in luminescent materials. By suppressing longitudinal optical phonon scattering at cryogenic temperatures, we effectively prevented exciton decoherence, transitioning the system into a globally delocalized state characterized by coherent radiative acceleration. This thermal and structural reconfiguration yields an extreme radiative lifetime of 13.11 ps alongside a high light yield of 21,851 photons MeV\textsuperscript{-1}. The resulting prompt photon brilliance translates directly into advanced diagnostic capabilities, as evidenced by the 30.8~ps coincidence time resolution and the precise temporal reconstruction of picosecond-scale electron bunches and gamma-ray pulses. Crucially, our findings confirm that the temporal bottleneck in extreme radiation detection has been definitively shifted from the intrinsic physical limits of scintillators to the bandwidth ceiling of current photoelectric readouts. Ultimately, this GOS-driven coherent emission paradigm establishes a robust theoretical and material framework for next-generation deep-picosecond detectors, promising profound impacts across high-luminosity colliders, advanced accelerator diagnostics, and precision medical imaging.

\section*{Acknowledgments}
We thank Z. Yuan for constructing the X-ray excited TCSPC system and the members of S. Huang's group for their assistance with the superconducting radio-frequency accelerator application tests.

\subsection*{Author Contributions} 
    \noindent \hspace{2em} Conceptualization: CD, YZ, XO, JX, XPO
    
    \noindent \hspace{2em} Methodology: YZ, CD, HH, YL, JB,YD, SW, JS, SH, WM
    
    \noindent \hspace{2em} Investigation: CD, YZ, HH, YL, YD, SW, JS, SH, WM
    
    \noindent \hspace{2em} Visualization: YZ, CD
    
    \noindent \hspace{2em} Funding acquisition: JX, YZ
    
    \noindent \hspace{2em} Project administration: JX, XPO, XO
    
    \noindent \hspace{2em} Supervision: JX, XPO, XO
    
    \noindent \hspace{2em} Writing -- original draft: YZ, CD
    
    \noindent \hspace{2em} Writing -- review \& editing: CD, YZ, XO, HH, YL, JS, YS, YD, WM, SH, SW, JX, XPO

\subsection*{Conflicts of Interest}
\textbf{ }\textbf{Y. Z., J. X., C. D., Y. S. and}\textbf{ J. }\textbf{B}\textbf{. }filed China National Intellectual Property Administration\textbf{ }provisional patent applications related to this work . The other authors declare that they have no\textbf{ }competing interests.

\subsection*{Data Availability}
All data are available in the main text or the supplementary materials.

\section*{Supplementary Materials}
    \noindent \hspace{2em} Materials and Methods 
    
    \noindent \hspace{2em} Supplementary Text 
    
    \noindent \hspace{2em} Figs. S1 to S20

\bibliographystyle{unsrtnat}
\bibliography{Manuscript}

\end{document}